# *Properties of a thin metal layer as a tool in focusing (projecting) procedure*

V.S.Zuev

The P.N.Lebedev Physical Institute
53 Leninsky pr., 119991 Moscow Russia

The field of an evanescent wave in a space with a thin metal layer is explored. The wave number of the evanescent wave exceeds many times the wave number of a uniform plane wave in adjoining gaps. In agreement with the results of the Pendry's paper of the 2000 year the field behind the metal layer occurs to be amplified as compared to the field in absence of the metal layer. Pendry has demonstrated the effect for the metal of the dielectric constant of -1. In this paper it is shown that the effect exists for the dielectric constant of -15 and for an arbitrary thickness of the adjoining gaps. This widens the number of possible applications. The metal losses do not destoy the effect appreciably.

# *Фокусирующие (проектирующие) свойства тонкого слоя металла*

В.С.Зуев

Физический ин-т им. П.Н.Лебедева РАН
119991 Москва, Ленинский пр-т 53

Рассмотрено поле исчезающей волны в пространстве с тонким слоем металла ($\varepsilon < 0$). Волновое число исчезающей волны заметно превышает $k_0 = 2\pi/\lambda_0$, волновое число однородной плоской волны в прилегающих к серебру промежутках. В согласии с результатом работы Pendry 2000-го года поле за металлическим слоем оказывается усиленным по сравнению с полем в отсутствие слоя металла. Pendry продемонстрировал эффект для $\varepsilon = -1$, а в данном рассмотрении показано, что эффект существует и для $\varepsilon \approx -15$ и при произвольных толщинах прилегающих промежутков. Это расширяет возможности применений явления. Установлено слабое влияние потерь в металле на качество изображения за слоем серебра.

Данная работа предпринята с целью изучения идеи Pendry /1/ о линзе субволнового разрешения. Автор показал, что можно построить изображение объекта с разрешением, которое превышает разрешение обычного объектива. Это возможно при применении гипотетических материалов с отрицательным показателем преломления $n = -1$. При их применении восстанавливается не только фаза распространяющихся волн, но и амплитуда исчезающих волн (evanescent waves). Если иметь в виду практически доступные материалы, то подобный эффект будет, как показано в /1/, наблюдаться в опытах с тонкой пленкой серебра на частоте плазмонов с бесконечно большим волновым числом (так называемые локализованные плазмоны), которые существуют в пленках серебра на частоте, где $\varepsilon = -1$. Отличия от идеальности будут возникать вследствие материальных потерь в металле.

Ниже рассмотрены как слои металла с $\varepsilon = -1$, так и с $\varepsilon \approx -15$. Как оказалось, восстановление амплитуды исчезающих волн происходит и при $\varepsilon \approx -15$, то есть существование эффекта не ограничено условием $\varepsilon = -1$.

Работа /1/ инициировала многочисленные последующие работы. Рассматривались такие постановки опыта, когда $\varepsilon_{met}$ равно с обратным знаком $\varepsilon_{gap}$ вещества в прилегающих промежутках (по крайней мере, в одном из промежутков). При этом относительная диэлектрическая проницаемость металла оказывается равной $-1$. Отметим работу группы Pendry /2/, где в деталях изучен эффект металлического слоя в симметричном и несимметричном окружении с учетом потерь и волнового запаздывания. В работе /3/ экспериментально отчетливо продемонстрирован эффект увеличения качества мелких деталей в изображении в опыте с плоским слоем серебра. С помощью работы /4/ можно найти то, что было опубликовано по данному вопросу ранее.

Предлагаемое ниже рассмотрение частично пересекается с рассмотрением в /2/, но в то же время сильно от него отличается. Рассматривается тонкий плоский слой металла в вакууме, см. рис.1, $\varepsilon_{1,3} = 1$, $\varepsilon_2 \leq -1$. В отличие от /2/ в расчете не предполагаются равенства $\varepsilon_2 = -\varepsilon_1$ или $\varepsilon_2 = -\varepsilon_3$; при этом допускается возможность $|\varepsilon_2| \gg 1$. Зазоры до слоя серебра и после него не полагаются равными половине толщины слоя серебра, как в /2/, а произвольны по толщине. Щели в объектной плоскости освещаются исчезающей волной с волновым числом $k_x > k_0 = \frac{\omega}{c}\sqrt{\varepsilon_1 \mu_1}$, а не плоской волной с $k_x = 0$. Рассмотрение не требует соблюдения условия квазистатичности.

Освещение объектных щелей исчезающей волной практически более целесообразно, чем освещение плоской поперечной волной, нормально падающей на плоскость, содержащей щели. В самом деле, два тонких слоя металла в каком-либо наноустройстве, каждый поддерживающий распространение поверхностных плазмонов, будут при необходимости взаимодействовать друг с другом скорее с помощью исчезающей волны, а не с помощью распространяющейся волны. Это вполне оправдывает предлагаемую ниже постановку задачи.

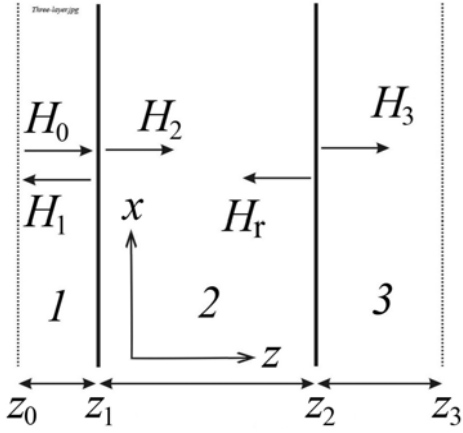
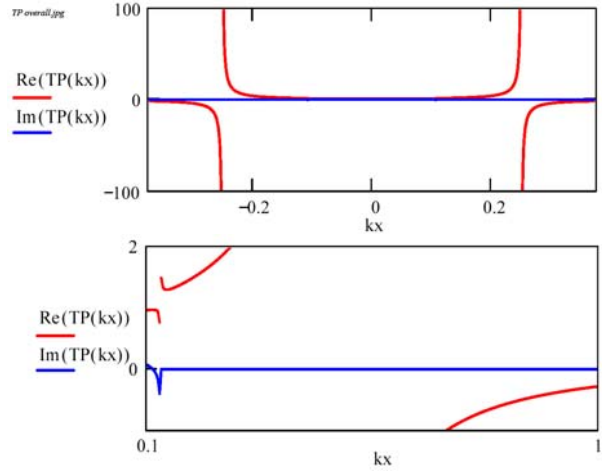

Рис.1. 1 и 3 – промежутки без серебра, 2 – слой $Ag$. Волны $H_i$ поляризованы вдоль оси $y$.

Рис.2. Функция пропускания $T_P(k_x)$ для слоя серебра толщиной $d = 6$ нм на длине волны 585 нм, $\varepsilon_2 = -14.882$. На верхнем фрагменте ось абсцисс имеет линейную шкалу, на нижнем фрагменте - логарифмическую шкалу.

Начало координат поместим на плоскость раздела сред 1 и 2. Будем считать, что последовательные плоскости слева направо имеют координаты $z_0$, $z_1$, $z_2$ и $z_3$. В объектной плоскости $z_0 < 0$, параллельной слою металла, имеется источник в виде двух параллельных щелей. Картина поля в объектной плоскости имеет вид:

$$\vec{H}_0 = \vec{a}_y H_{0y} e^{-\kappa_{0z} z} \cdot f(x),$$
$$f(x) = \cos k_{0x} x \cdot \{\exp[-(x-x_0)^m / \Delta x_0^m] + \exp[-(x+x_0)^m / \Delta x_0^m]\}, \quad (1)$$
$$m = 2^n, \ n = 1,2..., \ k_{0x}^2 - \kappa_{0z}^2 = k_0^2 = (\omega/c)^2 \varepsilon_1 \mu_1.$$

Рассматриваются поперечно-магнитные волны, поляризованные по оси $y$. Суммарная толщина промежутков $z_0 z_1$ и $z_2 z_3$ вне серебра равна $L$. Пропускание слоя серебра описывается функцией $T_P(k_x)$ /5/:

$$T_P(k_x) = \frac{H(z=z_2)}{H(z=z_1)} = \frac{4K}{(K+1)^2 e^{-\kappa_{2z} d} - (K-1)^2 e^{\kappa_{2z} d}}, \quad (2)$$
$$\kappa_{1z} = \sqrt{k_x^2 - (\omega_0/c)^2 \varepsilon_1 \mu_1}, \ \kappa_{2z} = \sqrt{k_x^2 - (\omega_0/c)^2 \varepsilon_2 \mu_2}, \ K = -\frac{\kappa_{2z} \varepsilon_1}{\kappa_{1z} \varepsilon_2}.$$

Это выражение совпадает с выражением для $T_p(k_x)$ из /2/, если в последнем положить $k_z^{(1)} = i\kappa_{1z}$, $k_z^{(2)} = i\kappa_{2z}$, $\varepsilon_1 = \varepsilon_3$ и $\mu_1 = \mu_2 = \mu_3 = 1$. Ниже принято, что все магнитные проницаемости $\mu_i$ равны 1. Функция $T_P(k_x)$ для слоя серебра толщиной 6 нм на длине волны 585 нм, $\varepsilon_2 = -14.882$, приведена на рис.2. Кривая имеет особенности при значениях $k_x$, равных волновым числам двух резонансных поверхностных плазмонов, симметричного $k_s = 1.075 \cdot 10^5 \, см^{-1}$, и антисимметричного $k_a = 2.498 \cdot 10^5 \, см^{-1}$.

Поле по формуле (1) и спектр волновых чисел $Gt(k_x) = \frac{1}{\sqrt{2\pi}} \int_{-\infty}^{\infty} f(x) e^{-ik_x x} dx$ составляющих его исчезающих волн $e^{ik_x x - \kappa_{1z} z}$ изображены на рис.3. На отрезке от $z_0$ до $z_3$ компоненты поля $e^{ik_x x - \kappa_{1z} z}$ подвергаются трансформации по формуле

$$GG(k_x) = Gt(k_x) T_P(k_x) e^{-\kappa_{1z}(k_x) L}. \tag{3}$$

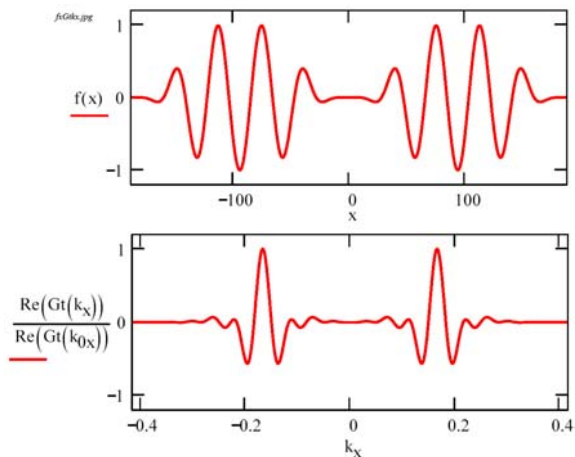
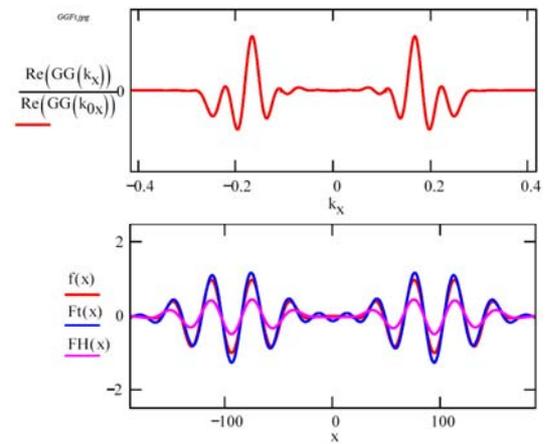

Рис.3. Значение 100 на оси $x$ соответствует 1000 нм, значение 0.2 по оси $k_x$ соответствуют $2 \cdot 10^5 \, см^{-1}$.

Рис.4. Описание – в тексте.

Волновое число $k_{0x}$ выбрано в промежутке между $k_s$ и $k_a$. Оно равно $k_a / 1.5 = 1.67 \cdot 10^5 \, см^{-1}$, длина волны $\lambda_0 = 376.2 \, нм$. Расстояние между серединами щелей $2x_0 = 2 \cdot 2.5 \cdot \lambda_0 = 1881 \, нм$, полная ширина каждой из щелей $2\Delta x_0 = 3 \cdot \lambda_0 = 1129 \, нм$. Толщина слоя серебра $d = 6 \, нм$. Суммарная толщина $L$ промежутков вне слоя серебра равна $10d = 60 \, нм$. Вычисленный по формуле (3) спектр изображен на рис.4, верхний фрагмент. Из сравнения с кривой на рис.3, нижний фрагмент, видны возникшие искажения при больших значениях $|k_x|$. Поле в плоскости $z_3$ получаем по формуле обратного фурье-преобразования

$$Ft(x) = \frac{1}{\sqrt{2\pi}} \int\limits_{0.7k_{0x}}^{1.3k_{0x}} Gt(k_x)TP(k_x)e^{-q1z(k_x)L}(e^{ik_xx} + e^{-ik_xx})dk_x. \qquad (4)$$

Интегрирование в (4) проделывается в несколько ограниченном интервале значений $k_x$, что означает фильтрование. Результат вычислений по формуле (4) представлен синей кривой на нижнем фрагменте рис.4. Это поле в плоскости изображения $z_3$. Поле в плоскости $z_0$ на этом рисунке изображено красной кривой. Пурпурным цветом изображено поле в плоскости $z_3$ в отсутствие слоя серебра. Для этого случая кривая рассчитана по формуле

$$FH(x) = \frac{1}{\sqrt{2\pi}} \int\limits_{0.7k_{0x}}^{1.3k_{0x}} Gt(k_x)e^{-q1z(k_x)(L+d)}(e^{ik_xx} + e^{-ik_xx})dk_x. \qquad (5)$$

Видно, что поле в плоскости $z_3$ в опыте со слоем серебра довольно хорошо совпадает с полем в объектной плоскости $z_0$. Наблюдаются некоторые искажения, но они невелики. Отметим, что плоскость $z_3$ находится на удалении $L + d = 66\,нм$ от объектной плоскости. В опыте без слоя серебра поле в этой плоскости оказывается двукратно ослабленным, пурпурная кривая.

Теперь обсудим случай близких по абсолютной величине значений $\varepsilon_1$ и $\varepsilon_2$. На длине волны $430.8\,нм$ для серебра $\varepsilon_2 = -6.06 + i0.197$ /6/, для алмаза - $\varepsilon_1 = 6.042$ /7/, отношение $\varepsilon_2/\varepsilon_1$ равно $-1.003$. Ниже в расчете отношение $\varepsilon_2/\varepsilon_1$ выбрано равным $-1.01$. Для $\varepsilon_3$ выбираем значение, равное $\varepsilon_1$.

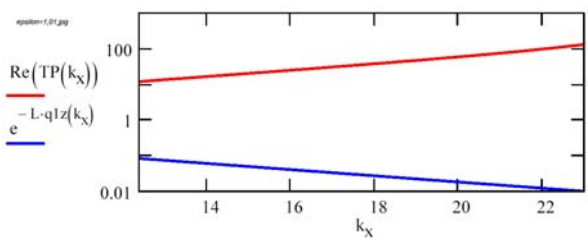 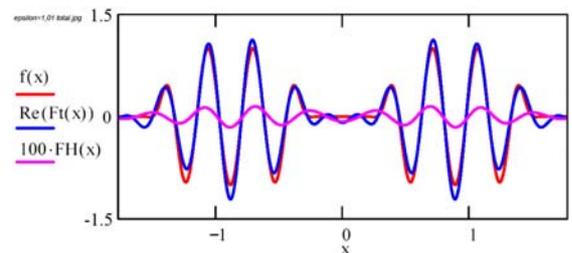

Рис.5.   Рис.6.

Первоначально считаем $\operatorname{Im}\varepsilon_2 = 0$. Толщина слоя серебра $d = 2\,нм$. Суммарная толщина $L$ промежутков вне слоя серебра равна $d$. Соответствующие волновые числа равны: $k_0 = 1.458 \cdot 10^5\,см^{-1}$, $k_s = 1.459 \cdot 10^5 \, см^{-1}$ и $k_a = 2.653 \cdot 10^7 \, см^{-1}$. Волновое число $k_{0x}$ выбрано равным $(1/1.5)k_a = 1.769 \cdot 10^7 \, см^{-1}$. На рис.6 приведены функция $T_P(k_x)$, верхняя, красная кривая, и функция $e^{-\kappa_{1z}L}$, нижняя, синяя кривая, в интервале значений $k_x = (0.7 \div 1.3)k_{0x}$. По оси ординат масштаб логарифмический. Функция $e^{-\kappa_{1z}L}$ принимает значения от ~ 0.1 до 0.01. Соответственно квадрат этой функции, то есть пропускание слоя толщиной $L + d = 2d$ меняется в пределах от 0.01 до 0.0001. В опыте без слоя серебра сигнал в плоскости $z_3$ оказывается очень слабым. На рис.6 этот сигнал изображен 100-кратно увеличенным пурпурной кривой. Кроме того, видно, что сигнал сильно искажен. Промежуток между импульсами практически исчез.

Как и в предыдущем случае в опыте со слоем серебра наблюдается хорошая передача сигнала из плоскости $z_0$ в плоскость $z_3$. Исходный сигнал изображен красной кривой, итоговый сигнал изображен синей кривой. Оба сигнала мало отличаются. Наблюдаются небольшие искажения в виде пульсаций в промежутке между импульсами. Однако контраст довольно велик и, соответственно, пространственное разрешение можно признать хорошим.

Теперь учтем потери в металле. Для диэлектрической проницаемости металла выбрано значение $\varepsilon_2 = -6.06 + i0.197$, для прилегающего слоя - $\varepsilon_1 = 6.042$ /7/. Остальные параметры – как для рис.6. Результат расчета приведен на рис.7. Отличия заметны лишь на кривой $GG(k_x)$. Функция $Ft(x)$ с потерями, рис.7, практически не отличима от таковой без потерь, рис.6 слева. Таким образом, можно признать, что при выбранных значениях параметров влияние потерь незначительно. Этот результат отличается от результата Pendry, представленного в /2/.

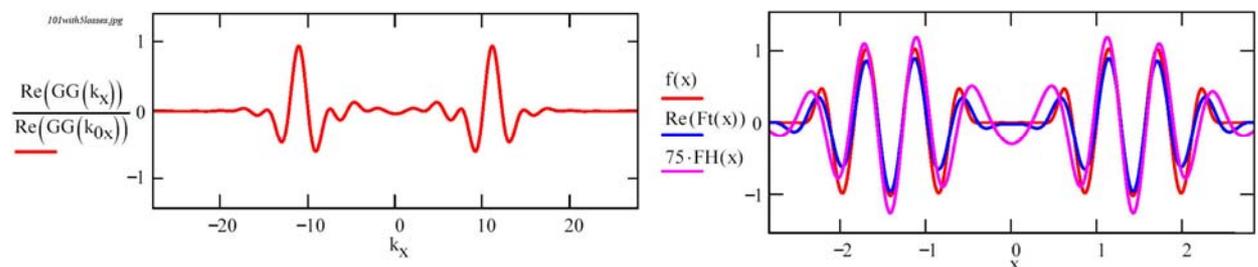

Рис.7. Для $\varepsilon_2 = -14.882 + i0.3858$. Остальные параметры – как для рис.4.

Выбранные выше значения толщины слоя серебра довольно малы в смысле допустимости выбора для $\varepsilon_2$ значений, определенных для массивных образцов серебра. Согласно /8/ в наночастицах (кластерах) атомов хороших металлов диаметра $2R = 10 \, нм$ длина свободного пробега электронов заметно сокращается из-за возрастающей роли столкновений с поверхностью. Для серебра длина свободного пробега $l_\infty$ электронов в массивном образце равна $52\,нм$ при $273\,K$ /8/. Уменьшение длины пробега приводит к изменению диэлектрической проницаемости. Наблюдается зависимость вида $1/R$ ширины плазмонного резонанса наночастицы при значениях $R \leq 6\,нм$ практически без изменения частоты резонанса. Для пленки толщиной $6\,нм$, по-видимому, следует брать $\operatorname{Im}\varepsilon_2$ увеличенным в $1.5-2$ раза в сравнении со значением для массивных образцов, для пленки $2\,нм$ рост $\operatorname{Im}\varepsilon_2$, по-видимому, во много раз больше. Поэтому в расчете, результат которого приведен на рис.7, значение $\operatorname{Im}\varepsilon_2$ выбрано увеличенным в 5 раз против значения $0.197$.

    Сконструируем устройство, в котором будет осуществляться проектирование пространственно-ограниченного поля из одной плоскости в другую, удаленную плоскость. Это устройство может состоять из двух параллельных полосковых диэлектрических волноводов. Щели на обращенных друг к другу границах этих волноводов будут осуществлять связь между волноводами. Введение слоя из серебра будет усиливать связь первого волновода со вторым и осуществлять пространственно определенное возбуждение второго волновода. Такую же связь можно осуществить и между двумя металлическими полосковыми волноводами, поддерживающими распространение поверхностных плазмонов. В сложном устройстве, состоящем из множества слоев, проектирование поля с помощью серебряного слоя позволит осуществить разветвленные, но селективные связи между слоями.

    Подведем итог. Тонкий слой металла с отрицательной диэлектрической проницаемостью $\varepsilon_2$ и малыми потерями усиливает поле исчезающей волны. В результате оказывается возможным воспроизвести с хорошей точностью пространственное распределение поля в плоскости перед слоем металла в плоскости за слоем металла. Эффект ранее рассматривался в /1/ при $\varepsilon_2 = -1$. Проделанное рассмотрение показывает, что восстановление амплитуды исчезающих волн происходит и при $\varepsilon \approx -15$ и при произвольных толщинах прилегающих промежутков. Установлено слабое влияние потерь в металле на качество изображения за слоем серебра. Это расширяет возможности применения явления. В сложном устройстве, состоящем из множества нанослоев, проектирование поля с помощью серебряного слоя позволит осуществить разветвленные, но селективные связи между слоями.